\documentclass{elsarticle}

\usepackage[english]{babel}

\usepackage[letterpaper,top=2cm,bottom=2cm,left=3cm,right=3cm,marginparwidth=1.75cm]{geometry}

\usepackage{subcaption}
\usepackage{float}

\usepackage{tabularx}

\usepackage{booktabs}
\usepackage{array}

\usepackage{amsmath}
\usepackage{graphicx}
\usepackage{xcolor}
\usepackage[colorlinks=true, allcolors=blue]{hyperref}

\usepackage{tikz}
\usetikzlibrary{arrows.meta, positioning, calc, shapes.geometric}

\begin{document}
\begin{frontmatter}

\title{Where should we meet? Ad hoc selection of emergency staging areas with open data} 

\author[label1,label2]{Moritz Schneider\corref{cor1}}

\cortext[cor1]{Corresponding author}

\author[label2]{Gustav Leinen}

\author[label2]{Alexander Fekete}

\author[label1]{Tina Comes}

\affiliation[label1]{
  organization={German Aerospace Center, Institute for the Protection of Terrestrial Infrastructures},
  city={Sankt Augustin},
  postcode={53757},
  country={Germany}
}

\affiliation[label2]{
  organization={Institute of Rescue Engineering and Civil Protection, TH Köln - University of Applied Sciences},
  city={Cologne},
  postcode={50678},
  country={Germany}
}

\begin{abstract}
Urban emergency operations often require temporary staging areas, where personnel, vehicles and equipment can be assembled and dispatched. Unlike shelters, warehouses or depots, such staging areas cannot be planned in advance because the locations of the hazard are unknown until they occur. Due to the multitude of competing requirements, ranging from safety to capacity and access, the rapid identification of suitable staging areas can overwhelm incident commanders and impede timely decision-making.
This paper introduces a spatial decision support system that translates requirements from practice into a reproducible ranking of sites.
The underlying criteria and their relative importance were derived from an online survey of more than 100 fire-brigade and civil-protection practitioners in Germany.
Eight spatially assessable criteria are translated into a model operating on openly available geographic data, by combining feasibility filters with a multi-criteria evaluation. 
We demonstrate the system by using the case of an unexploded ordnance (UXO) in Cologne, an operation that affected about 20,000 residents in June 2025. The system 
produces a short list of suitable staging areas that is robust to plausible changes of weights and thresholds, and the area actually used in the real operation is among its six highest-ranked candidates.
These results show how our approach can identify candidate sites and make transparent the rationales underlying their ranking. Our contribution is a practitioner-informed approach to emergency staging area selection.
\end{abstract}

\begin{keyword}
Disaster Risk Management \sep Selection Problem \sep GIS \sep Multi-criteria decision analysis
\end{keyword}

\end{frontmatter}

\section{Introduction}
Urban emergencies require responding organisations to assemble their forces somewhere close, but outside the area at risk. Gas leaks, fires, hazardous material releases, or the discovery of explosive ordnances all create such emergencies. 
Typically, staging areas are sizeable, readily accessible open spaces where the responding forces can be briefed, equipped, and dispatched, and where emergency vehicles, material, and personnel can be assembled and managed \citep{fwdv100,kapucu2007staging}. As such, the location of the staging areas connects the logistical preparations with areas, where assistance is required. Therefore, urban staging areas must provide sufficient space to accommodate all resources, vehicle access and good connections to deployment locations while remaining out of harm's way. Today, this selection relies predominantly on the individual experience and understanding of the commanding officer, in line with the recognition-based decision making \citep{klein2010rapid}. What is missing is a systematic and data-driven approach to rapidly and transparently assess candidate locations. \\

Staging-area selection is an incident-specific decision within the broader emergency logistics planning field \citep{vanajakumari2016integrated}. While preparedness activities can identify potential locations, their suitability must match the location and scale of an incident as well as the access conditions. Sites that appear suitable may also be occupied or not available for response operations.  

Research on data-driven site selection in humanitarian and emergency logistics has established methods to address a range of facility location problems, such as shelter placement \citep{kilci2015,ma2019shelter}, depot and logistics-centre siting \citep{balcik2008,feng2023}, and the siting of emergency operations centres \citep{alidadi2025}. All these problems have their roots in optimisation and operations research, and assume that the problem characteristics are know a priori \citep{MELO2009401}. However, staging-area locations cannot be pre-planned. What is needed is therefore a transparent approach for rapidly identifying and screening candidate staging areas once the incident site is known. This approach needs to distinguish hard constraints from preferences and make transparent the basis of any recommendation. 

To bridge this gap, we introduce a spatial decision support system (DSS) \citep{keenan2019} that translates operational requirements from practice into a computational application to identify suitable staging areas.
The DSS rests on two integrated components.
\begin{itemize}
    \item \textbf{Requirements from practice:}
          To ensure that the selection process reflects the priorities and constraints of practitioners, we establish a set of criteria for staging areas, together with their relative importance, derived from an online survey of fire-brigade and civil-protection personnel in Germany.
    \item \textbf{Computational model:}
          Each spatially assessable criterion is translated into a model that operates on open geographic data, which enables the automated and reproducible scoring of candidate staging areas, independent of the hazard location and urban context.
\end{itemize}
The criterion scores are combined via a multi-criteria decision analysis into an overall suitability ranking of the candidate areas, which serves as the basis of the final map-based decision support system.
The methodological building blocks of this combination -- hard feasibility filters, saturating utility functions, and additive weighting -- are deliberately established ones.

The contribution of this paper is in formulating an approach for ad-hoc staging area selection as a class of facility location problems. We provide a requirement set with relative priorities from more than 100 practitioners from Germany across the organisations that share staging areas. We then translate the requirements into a model that generates and ranks candidates from open data for any hazard location. To validate our results, we compare the outcome for a case study in Cologne with a real operational decision, showing that saturating utilities reduce the choice to feasibility followed by proximity. 

The remainder of this paper is organised as follows. First, the conceptual background of optimal site selection is outlined (Section~\ref{sec:background}).
Second, the online survey is presented, together with the stakeholder requirements it reveals for staging-area selection (Section~\ref{sec:survey}).
Third, the computational decision model is introduced, translating these requirements into a data-driven selection problem using utility functions and multi-criteria decision analysis (Section~\ref{sec:model}).
Fourth, a case study of an UXO evacuation in the city of Cologne, Germany, demonstrates how the decision support system supports staging-area selection (Section~\ref{sec:casestudy}).
Finally, the model and its results are discussed, and directions for future research are identified (Section~\ref{sec:discussion}).

\section{Background} \label{sec:background}
\subsection{Spatial site selection in emergency and humanitarian logistics}

The selection of suitable locations for facilities, resources, or operational areas is a long-standing problem at the intersection of operations research, geography, and decision science \citep{church1974,malczewski2006}.
In emergency management and humanitarian logistics, well-studied application areas include the location of shelters and camps for affected populations, the placement of storage facilities and humanitarian depots, the siting of fire stations and other emergency medical facilities, the planning of evacuation areas and routes, and the establishment of disaster-operation sites such as search-and-rescue bases, staging areas, and command posts \citep{balcik2008,caunhye2012}. Recent applications combine geographic layers with a multi-criteria weighting of the candidate locations for emergency logistics facility locations \citep{lin2026logistics}.
Although these location problems differ in scale, time horizon, and institutional setting, they share three of challenges in identifying the best possible location: uncertainty, conflicting priorities and accountability.\\

\textit{Uncertainty:} the location and magnitude of the demand, or of the hazard itself, are usually unknown or at least uncertain at the time of the decision. To address this, often robust optimisation approaches are chosen, whereby the chosen location must perform sufficiently well under a range of possible scenarios \citep{snyder2006,li2025robustlocation}. In addition, there is data uncertainty. 
The detailed geospatial information on infrastructure, land use, and the affected area, may also not be fully reliable or up to date \citep{haklay2010,Schneider2026}. As such,  it is also harder to assess whether a location respects the required hard constraints, such as minimum safety distances, land-use restrictions, or load-bearing requirements \citep{malczewski2006}.

\textit{Trade-offs between heterogeneous criteria:} safety, accessibility, proximity to the affected population, capacity, and compatibility with the surrounding environment often conflicts, so that no candidate location is unambiguously superior to the others \citep{belton2002}. Moreover, in emergency operations, several authorities with different responsibilities and priorities are e served by a single location decision, which makes the resolution of trade-offs even more challenging \citep{kapucu2006interagency}.

\textit{Accountability:} because site selection decisions in emergencies carry high operational and reputational stakes, they are expected to be transparent, explainable, and reproducible, thereby also increasing accountability for the decisions made.\\

With respect to the \textit{time horizon} of the decision, approaches fall into the preparedness or the response phase of disaster operations \citep{altay2006}.
Planned, \emph{ex-ante} facility location problems, such as the placement of storage facilities or fire stations, are decided before any specific event occurs and can be based on long-term demand estimates and strategic objectives.
Ad-hoc, \emph{ex-post} facility location problems, such as the establishment of evacuation areas or disaster-operation sites, must be decided after an event has occurred. Because they respond to urgent needs, these decisions are subject to tremendous time pressure.
Table~\ref{tab:background_sitesel} gives an overview of the application areas, the decision horizon they are usually associated with, and the criteria and methods that are typically used; it is intended as an orientation within the neighbouring fields rather than as a systematic review.

\begin{table}[htbp]
\centering
\caption{Overview of related spatial site selection problems in emergency and humanitarian logistics. The table is intended as an orientation within neighbouring application areas and does not claim to be a systematic review.}
\label{tab:background_sitesel}
\footnotesize
\setlength{\tabcolsep}{4pt}
\renewcommand{\arraystretch}{1.1}
\begin{tabularx}{\textwidth}{@{}>{\raggedright\arraybackslash}p{2.8cm} >{\raggedright\arraybackslash}p{1.8cm} >{\raggedright\arraybackslash}X >{\raggedright\arraybackslash}p{3.0cm} >{\raggedright\arraybackslash}p{1.1cm}@{}}
\toprule
\textbf{Application area} & \textbf{Decision horizon} & \textbf{Typical criteria} & \textbf{Predominant methods} & \textbf{Refs.} \\
\midrule
Shelter siting and allocation & ex ante and ex post & Capacity, accessibility, hazard exposure, coverage of the affected population & MCDA, GIS-based MCE, location optimisation & \citep{kilci2015,ma2019shelter,albuquerque2024shelters} \\
\addlinespace[2pt]
Storage facilities and humanitarian hubs & ex ante & Demand coverage, transport cost, connectivity to the transport network, risk exposure & Location optimisation, MCDA & \citep{balcik2008,feng2023,lin2026logistics} \\
\addlinespace[2pt]
Emergency service facility siting (fire, EMS) & ex ante & Response time, population coverage, equity of service provision & Location optimisation, GIS-based MCE & \citep{toregas1971,erden2010} \\
\addlinespace[2pt]
Evacuation areas and routes & ex post & Distance and travel time to the hazard, road capacity, safety of the receiving area & Network analysis, GIS-based MCE, location optimisation & \citep{cova1997,alcada2009,moradi2025evacuation} \\
\addlinespace[2pt]
Disaster-operation sites (Command posts, staging areas) & ex post & Available area, accessibility, safety distance, supply and support infrastructure & MCDA, GIS-based MCE & \citep{kapucu2007staging,alidadi2025} \\
\addlinespace[2pt]
\textbf{Ad-hoc staging areas around an exclusion zone (this work)} & ex post, time-critical & Hard safety distance, area and manoeuvring space, travel time to the affected area, connection to main routes and fuel supply & GIS-based MCE with survey-based weights & --- \\
\bottomrule
\end{tabularx}
\end{table}

\subsection{The ad hoc staging area selection problem}
Compared with the application areas above, staging area selection shifts the emphasis from a priori facility location optimisation to an ad hoc operational decision tied to a specific incident. The selection of staging areas combines four features: (i) the decision is taken ad hoc and under time-pressure, since the hazard location or access constraints are not known a priori. (ii) Safety constraints such as the required distance to the hazard-affected area exclude parts of the city before any ranking. (iii) The candidate solutions have to be identified from urban land-use data; there is no pre-compiled list. (iv) The relative importance of the criteria should reflect the operational practice of the organisations that share and use the site, which motivates the survey presented in the following section.

The contribution of this work therefore lies in rapid decision-support and problem elicitation: the methodological effort lies in the elicitation and the spatial modelling of the criteria rather than in the aggregation, which is deliberately kept simple and traceable, since a ranking that cannot be explained at a command post will not be used there.

\section{Practitioner Survey on Staging Area Requirements}
\label{sec:survey}
\begin{figure}[htbp]
  \centering
  \includegraphics[width=0.9\linewidth]{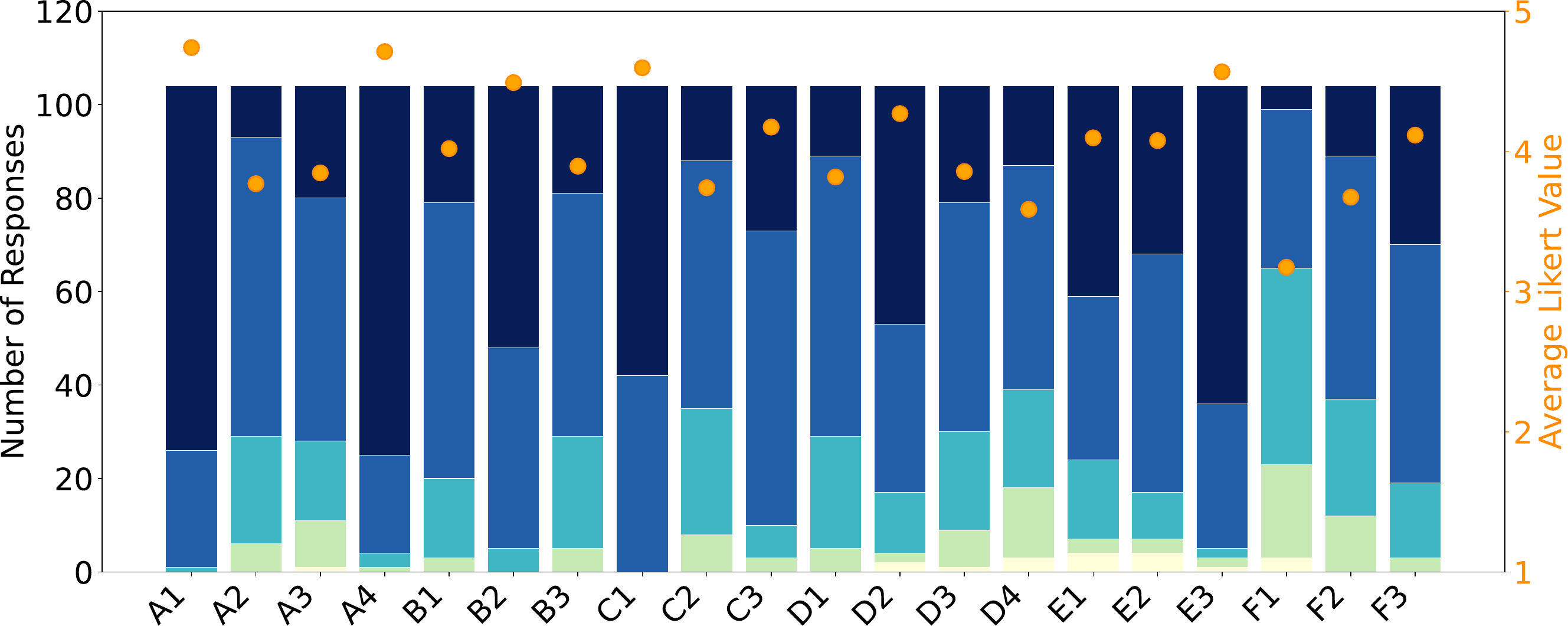}
  \caption{Distribution of the Likert ratings per requirement: 1 = very irrelevant (light shading), 5 = very important (dark shading).}
  \label{fig:likert}
\end{figure}

Based on a review of grey literature, in particular German guidelines in the context of staging areas \citep{fwdv100}, we developed a list of 20 requirements, grouped into six categories (see Table~\ref{tab:questions}): \emph{Location and Safety}, \emph{Traffic Accessibility}, \emph{Area Provision and Capacity}, \emph{Infrastructure and Supply}, \emph{Leadership and Communication}, and \emph{Logistics and Personnel}.
For each statement, participants indicated its importance on a 5-point Likert scale ranging from 1 (\emph{highly irrelevant}) to 5 (\emph{very important}); the distribution of these ratings is shown in Figure~\ref{fig:likert}.
In addition, the six categories were ranked against each other, with rank~1 corresponding to the most important category.
The forced ranking of the categories is therefore an essential complement to the Likert ratings, as it enforces a prioritisation that the ratings alone, due to their ceiling effect, cannot provide; for this reason the category ranking, and not the per-requirement ratings, is used to derive the criterion weights of the selection model (Section~\ref{sec:utility}).
\\

\begin{table}[h]
  \caption{Survey questions regarding requirements on staging areas.}
  \label{tab:questions}
  \centering
  \small
  \begin{tabular}{l l l}
    \textbf{Enum} & \textbf{Category} & \textbf{Question regarding} \\ \hline
    A1 & Location and Safety & Location outside the danger zone \\
    A2 & Location and Safety & Proximity to incident site \\
    A3 & Location and Safety & Psychological relief for personnel (visual separation) \\
    A4 & Location and Safety & Protection against secondary hazards \\
    B1 & Traffic Accessibility & Sufficient manoeuvring space without reversing \\
    B2 & Traffic Accessibility & Driveability for heavy vehicles \\
    B3 & Traffic Accessibility & Good connection to main traffic routes \\
    C1 & Area Provision and Capacity & Sufficient area for emergency vehicles \\
    C2 & Area Provision and Capacity & Separate entry and exit routes \\
    C3 & Area Provision and Capacity & Clear structure (vehicles, personnel, logistics) \\
    D1 & Infrastructure and Supply & Lighting of traffic and work areas \\
    D2 & Infrastructure and Supply & Sanitary facilities \\
    D3 & Infrastructure and Supply & Power supply (grid or emergency generator) \\
    D4 & Infrastructure and Supply & Water supply \\
    E1 & Leadership and Communication & Documentation and situation awareness capability \\
    E2 & Leadership and Communication & Setup capability for staging area management \\
    E3 & Leadership and Communication & Stable radio and communication connection \\
    F1 & Logistics and Personnel & Proximity to petrol station \\
    F2 & Logistics and Personnel & Rest and regeneration areas \\
    F3 & Logistics and Personnel & Catering options \\
    \hline
  \end{tabular}
\end{table}

In addition to the requirements themselves, the survey elicited background information on the participants' organisational affiliation and their practical experience with staging areas. Since staging areas are commonly shared by several response organisations, the survey targeted a broad range of stakeholder groups rather than a single organisation. The survey was distributed via mailing lists of several relevant organisations and individual distributors within the organisations; participation was voluntary, and all responses were collected anonymously.

In total, 307 questionnaires were returned, of which 104 (33.9\%) were complete and included in the analysis (Table~\ref{tab:organisation} shows an overview of the background of all participants).
Respondents were predominantly affiliated with organisations that regularly operate staging areas in real-world incidents: 60.6\% were from fire departments, 20.2\% from technical relief agencies, 39.4\% from aid organisations, and 15.4\% from municipal, district, or state administrations.
The majority of participants also reported direct experience: 80.8\% had deployed personnel in a staging area, 30.8\% had led a staging area in real operations, and 40.4\% in exercises, while 46.2\% were involved in conceptual-level planning.
Furthermore, 69.2\% of the participants reported 5--20 years of professional experience in civil protection, and 17.3\% more than 20 years. \\

\begin{table}[htbp]
  \centering
  \caption{Background of the 104 participants. Multiple answers were possible for organisation and experience, so the percentages of these blocks do not sum to 100\,\%.}
  \label{tab:organisation}
  \small
  \begin{tabular}{lcc}
    \toprule
    \textbf{Question} & \textbf{\%} & \textbf{n} \\
    \midrule
    \multicolumn{3}{l}{\textbf{Organisation}} \\
    \addlinespace[2pt]
    Fire Department & 60.6 & 63 \\
    Technical Relief Agency & 20.2 & 21 \\
    Aid Organisation & 39.4 & 41 \\
    Municipality / District / State & 15.4 & 16 \\
    Other & 2.9 & 3 \\
    \midrule
    \multicolumn{3}{l}{\textbf{Experience}} \\
    \addlinespace[2pt]
    Deployed personnel in a staging area & 80.8 & 84 \\
    Leading a staging area & 30.8 & 32 \\
    Leading a staging area in exercises & 40.4 & 42 \\
    Planning at conceptual level & 46.2 & 48 \\
    Support / Partial role & 28.8 & 30 \\
    \midrule
    \multicolumn{3}{l}{\textbf{Professional Experience in Civil Protection}} \\
    \addlinespace[2pt]
    $<$ 5 years & 13.5 & 14 \\
    5--10 years & 34.6 & 36 \\
    11--20 years & 34.6 & 36 \\
    $>$ 20 years & 17.3 & 18 \\
    \bottomrule
  \end{tabular}
\end{table}

The category ranking (see Table~\ref{tab:ranking}), based on the mean ranking, shows a clear hierarchy of relevance.
\emph{Location and Safety} was considered the most important category, with a mean rank of $\bar{r}=1.67$ and as first choice of 72 out of 104 respondents.
\emph{Area Provision and Capacity} ($\bar{r}=2.97$) and \emph{Leadership and Communication} ($\bar{r}=3.27$) follow as the second and third most important categories.
The remaining categories, \emph{Infrastructure and Supply} ($\bar{r}=4.20$), \emph{Traffic Accessibility} ($\bar{r}=4.40$), and \emph{Logistics and Personnel} ($\bar{r}=4.48$), were rated comparatively less important; their mean ranks are, however, only marginally different, so their relative ordering should be interpreted with caution.

The per-requirement Likert ratings (see Figure~\ref{fig:likert}) indicate that all requirements are perceived as at least somewhat important, since for the vast majority of requirements the majority of responses falls in the \emph{important} or \emph{very important} range (ratings 4--5).
Overall, the survey shows that the  safety of a staging area, as well as the provision of sufficient and clearly structured area are regarded as the most critical prerequisites, whereas infrastructure and logistical aspects, play a secondary role.

\begin{table}[htbp]
  \centering
  \caption{Ranking $\text{r}$ of categories including the amount of a category ranked as top category $n(\text{r}=1)$ and mean ranking $\bar{r}$.}
  \label{tab:ranking}
  \small
  \begin{tabular}{llcc}
    \toprule
    \textbf{Enum} & \textbf{Category} & \textbf{$n(\text{r}=1)$} & \textbf{$\bar{r}$} \\
    \midrule
    A & Location and Safety & 72 & 1.67 \\
    C & Area Provision and Capacity & 13 & 2.97 \\
    E & Leadership and Communication & 12 & 3.27 \\
    D & Infrastructure and Supply & 3 & 4.20 \\
    B & Traffic Accessibility & 0 & 4.40 \\
    F & Logistics and Personnel & 4 & 4.48 \\
    \bottomrule
  \end{tabular}
\end{table}

\subsection{Selection of Requirements for the Spatial Model}
\label{sec:requirementselection}
In this work, the requirements of the survey are mapped as far as possible onto spatial criteria. The computational model proposed here considers those survey requirements that admit such a formulation and uses them for staging-area selection. The 20 survey requirements were reviewed with respect to their suitability for a spatial selection process.
Requirements that describe the organisation or the equipment of a staging area rather than the suitability of a location were excluded, as they are relevant for operating a staging area but do not affect whether a site is selected in the first place. 
This applies, for instance, to rest and regeneration areas (F2) and catering options (F3), which can be provided on site, as well as to the setup of management and documentation capabilities (E1, E2) and the requirement of a clear structure of the site (C3).
Furthermore, the visual separation of the staging area from the incident site (A3) is highly dependent on the scenario considered -- in the case study presented in this work it is of no particular relevance as no destruction of the surrounding area is assumed, and protection against secondary hazards (A4) is already accounted for by the exclusion of the danger zone (A1).
Finally, a stable radio and communication connection (E3) cannot be derived reliably from static geographic data, since its coverage depends on terrain-dependent propagation and on the radio infrastructure in place at the time of the operation.

This selection is based on spatial assessability rather than on the surveyed importance: all traffic accessibility criteria (B1--B3) are included, as they are directly expressible via the road network and terrain data, whereas the infrastructure requirements of category D describe provisions to be installed at the site and hence lie outside the selection process.
In contrast, the proximity to a petrol station (F1) is assessable from point-of-interest data, since fuel infrastructure has a fixed, known location.
This selection yields the eight spatially assessable requirements A1, A2, B1, B2, B3, C1, C2, and F1, which the following subsection translates into spatially explicit models.

\section{Staging Area Selection Model}
\label{sec:model}
\begin{figure}[htbp]
\centering
\begin{tikzpicture}[
  font=\small,
  every node/.style={inner sep=0pt},
  blocktitle/.style={font=\small\bfseries, align=center, text width=3.9cm},
  blocktext/.style={font=\scriptsize, align=center, text width=3.9cm, text=black!75},
  flow/.style={-{Latex[length=2.8mm,width=2.0mm]}, line width=1.1pt, black!70, rounded corners=4pt},
  feed/.style={-{Latex[length=2.2mm,width=1.6mm]}, line width=0.7pt, black!50, dashed},
  frame/.style={draw=black!45, rounded corners=4pt, fill=black!3, line width=0.6pt},
  pics/picdata/.style={code={
      \draw[draw=black!35, fill=white] (-0.95,-0.55) rectangle (0.95,0.55);
      \fill[green!45!black!20, draw=black!35] (-0.80,-0.40) -- (-0.20,-0.45) -- (-0.12,0.02) -- (-0.72,0.10) -- cycle;
      \fill[green!45!black!20, draw=black!35] (0.20,0.12) rectangle (0.72,0.42);
      \draw[black!45, line width=0.6pt] (-0.95,0.28) -- (0.95,0.18);
      \draw[black!45, line width=0.6pt] (0.05,-0.55) -- (0.22,0.55);
      \fill[red!75!black] (0.45,-0.28) circle (0.10);
  }},
  pics/picmetric/.style={code={
      \fill[blue!35!black!12, draw=black!45] (-0.90,-0.42) -- (-0.25,-0.55) -- (0.02,0.08) -- (-0.45,0.48) -- (-0.95,0.15) -- cycle;
      \draw[black!55, dashed] (-0.48,-0.02) circle (0.30);
      \draw[{Latex[length=1.3mm]}-{Latex[length=1.3mm]}, black!70] (-0.78,-0.02) -- (-0.18,-0.02);
      \draw[black!55, line width=0.7pt] (0.35,-0.45) .. controls (0.60,-0.15) and (0.30,0.05) .. (0.55,0.45);
      \fill[black!65] (0.35,-0.45) circle (0.07);
      \fill[black!65] (0.55,0.45) circle (0.07);
  }},
  pics/picfilter/.style={code={
      \draw[red!65!black, line width=0.8pt] (-0.62,0.42) -- (-0.44,0.60)  (-0.44,0.42) -- (-0.62,0.60);
      \draw[red!65!black, line width=0.8pt] (0.44,0.42) -- (0.62,0.60)  (0.62,0.42) -- (0.44,0.60);
      \fill[black!8, draw=black!50] (-0.75,0.28) -- (0.75,0.28) -- (0.14,-0.14) -- (0.14,-0.48) -- (-0.14,-0.48) -- (-0.14,-0.14) -- cycle;
      \draw[green!45!black, line width=0.9pt] (0.30,-0.40) -- (0.44,-0.55) -- (0.72,-0.20);
  }},
  pics/picutility/.style={code={
      \draw[black!55, -{Latex[length=1.2mm]}] (-0.80,-0.45) -- (0.85,-0.45);
      \draw[black!55, -{Latex[length=1.2mm]}] (-0.80,-0.45) -- (-0.80,0.58);
      \draw[black!30, dashed] (-0.80,0.40) -- (0.80,0.40);
      \draw[blue!55!black, line width=0.9pt, domain=0:1.55, samples=40, variable=\x]
            plot ({-0.80+\x}, {-0.45+0.85*(1-(1-\x/1.55)*(1-\x/1.55))});
  }},
  pics/picmcda/.style={code={
      \fill[blue!60!black!75] (-0.85,0.26) rectangle (0.20,0.46);
      \fill[blue!60!black!55] (-0.85,-0.04) rectangle (-0.10,0.16);
      \fill[blue!60!black!40] (-0.85,-0.34) rectangle (-0.28,-0.14);
      \fill[blue!60!black!30] (-0.85,-0.64) rectangle (-0.34,-0.44);
      \node[font=\small, text=black!70] at (0.62,-0.10) {$\Sigma$};
      \draw[black!45] (0.30,0.50) -- (0.30,-0.68);
  }},
  pics/picrank/.style={code={
      \draw[draw=black!35, fill=white] (-0.95,-0.55) rectangle (0.95,0.55);
      \draw[black!30, line width=0.6pt] (-0.95,0.10) -- (0.95,0.02);
      \fill[orange!80!red] (-0.42,0.06) circle (0.19);
      \node[font=\tiny, text=white] at (-0.42,0.06) {1};
      \fill[black!65] (0.18,0.34) circle (0.17);
      \node[font=\tiny, text=white] at (0.18,0.34) {2};
      \fill[black!65] (0.40,-0.28) circle (0.17);
      \node[font=\tiny, text=white] at (0.40,-0.28) {3};
  }},
  pics/picsurvey/.style={code={
      \fill[blue!15] (-0.95,-0.14) rectangle (-0.35,0.14);
      \fill[blue!40] (-0.35,-0.14) rectangle (0.20,0.14);
      \fill[blue!65] (0.20,-0.14) rectangle (0.95,0.14);
      \draw[black!45] (-0.95,-0.14) rectangle (0.95,0.14);
  }},
]

\def\bw{2.35}   
\def\bh{1.75}   

\draw[frame] (-\bw,-\bh) rectangle (\bw,\bh);
\pic at (0,0.85) {picdata};
\node[blocktitle] at (0,-0.10) {Input data};
\node[blocktext]  at (0,-0.95) {Candidate open areas, road graph, POIs, land surface, incident location (Sec.~\ref{sec:spatialmodels})};

\begin{scope}[shift={(5.2,0)}]
  \draw[frame] (-\bw,-\bh) rectangle (\bw,\bh);
  \pic at (0,0.85) {picmetric};
  \node[blocktitle] at (0,-0.10) {Spatial models};
  \node[blocktext]  at (0,-0.95) {Area, inner width, surface class, access roads, network travel times (Sec.~\ref{sec:spatialmodels})};
\end{scope}

\begin{scope}[shift={(10.4,0)}]
  \draw[frame] (-\bw,-\bh) rectangle (\bw,\bh);
  \pic at (0,0.85) {picfilter};
  \node[blocktitle] at (0,-0.10) {Hard constrains};
  \node[blocktext]  at (0,-0.95) {A1, B1, B2, C2 and the bounds of A2, C1 $\rightarrow$ feasible set (Sec.~\ref{sec:spatialmodels})};
\end{scope}

\begin{scope}[shift={(0,-4.6)}]
  \draw[frame] (-\bw,-\bh) rectangle (\bw,\bh);
  \pic at (0,0.85) {picutility};
  \node[blocktitle] at (0,-0.10) {Soft criteria};
  \node[blocktext]  at (0,-0.95) {A2, B3, C1, F1 mapped to utilities $U\in[0,1]$ by non-linear functions (Sec.~\ref{sec:utility})};
\end{scope}

\begin{scope}[shift={(5.2,-4.6)}]
  \draw[frame] (-\bw,-\bh) rectangle (\bw,\bh);
  \pic at (0,0.85) {picmcda};
  \node[blocktitle] at (0,-0.10) {Aggregation};
  \node[blocktext]  at (0,-0.95) {Weighted sum $U_{\mathrm{total}}=\sum_i w_i U_i$ of the four utilities (Sec.~\ref{sec:utility})};
\end{scope}

\begin{scope}[shift={(10.4,-4.6)}]
  \draw[frame] (-\bw,-\bh) rectangle (\bw,\bh);
  \pic at (0,0.85) {picrank};
  \node[blocktitle] at (0,-0.10) {Ranked top-$N$ map};
  \node[blocktext]  at (0,-0.95) {Decision support for the incident commander (e.g. see Sec.~\ref{sec:casestudy})};
\end{scope}

\begin{scope}[shift={(5.2,-7.75)}]
  \draw[frame, fill=blue!3] (-7.55,-0.85) rectangle (7.55,0.85);
  \pic at (-6.60,0.0) {picsurvey};
  \node[font=\small\bfseries, align=left, anchor=north west, text width=11.6cm] at (-5.40,0.62)
       {Practitioner survey (Sec.~\ref{sec:survey})};
  \node[font=\scriptsize, align=left, anchor=north west, text width=11.6cm, text=black!75] at (-5.40,0.12)
       {20 requirements in 6 categories $\rightarrow$ spatially assessable criteria (Sec.~\ref{sec:requirementselection});\\ category ranking $\rightarrow$ criterion weights $w_i$ (Eq.~\ref{eq:weights})};
\end{scope}

\draw[flow] (\bw,0) -- (5.2-\bw,0);
\draw[flow] (5.2+\bw,0) -- (10.4-\bw,0);
\draw[flow] (10.4,-\bh) -- (10.4,-2.30) -- (0,-2.30) -- (0,-4.6+\bh);
\draw[flow] (\bw,-4.6) -- (5.2-\bw,-4.6);
\draw[flow] (5.2+\bw,-4.6) -- (10.4-\bw,-4.6);

\draw[feed] (0,-7.75+0.85) -- (0,-4.6-\bh);
\draw[feed] (5.2,-7.75+0.85) -- (5.2,-4.6-\bh);

\end{tikzpicture}
\caption{Overview of the staging area selection model. The practitioner survey determines which requirements enter the model and how strongly they are weighted; the spatial models translate each criterion into a measurable quantity; hard constrains reduce the candidate set to the feasible sites, and the utilities of the soft criteria are aggregated into a single ranking that is handed to the incident commander as a map.}
\label{fig:method_overview}
\end{figure}
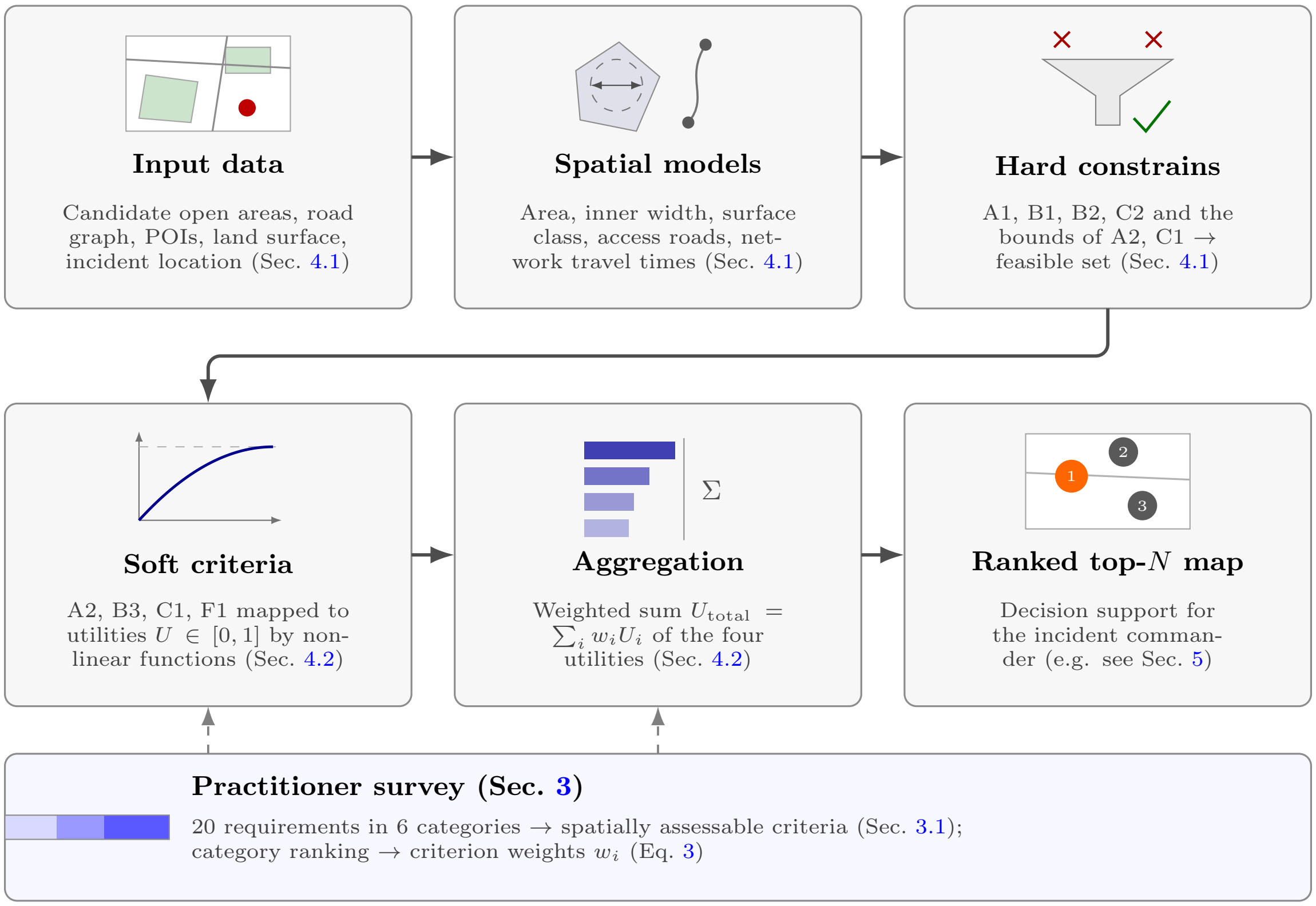

The requirements established in Section~\ref{sec:survey} are translated into a computational selection model that identifies and ranks suitable staging areas for a given incident and results in a map showing the top-$N$ staging areas (see Figure~\ref{fig:method_overview} for an overview of the full method and respective sections).
The model operates on five spatial data sets that are extracted from OpenStreetMap (OSM): 
(i) open areas (e.g., parking lots and brownfield sites) that serve as candidates for staging areas; (ii) the road network represented as a directed graph with edge travel times derived from the speed limits for emergency vehicles of the respective road classes (see \citep{Priesmeier2026}); (iii) a layer comprising of points of interest (POIs), i.e. places within the evacuation area at which the workload of responding forces are expected to be higher during the evacuation, such as schools, care facilities, tourist attractions, and train stations (similar to \citep{Schneider2026}); (iv) the locations of fuel stations; and (v) a land surface layer describing the ground cover. \\

The staging area selection proceeds in four processing steps (given the input data and resulting in the ranked top-$N$ map).
First, a set of \textit{spatial models} assigns to each candidate site the metrics required to assess these criteria (Section~\ref{sec:spatialmodels}).
Second, the criteria are applied either as \emph{hard} constraints that lead to immediate exclusion or as criteria whose partial fulfilment is expressed through utility functions (Section~\ref{sec:utility}). 
Hard constraints express binary feasibility conditions: a candidate site either satisfies the requirement or is excluded from further consideration. 
Criteria, in contrast, can have a minimum or maximum value that determines feasibility, but within this scope, the degree of fulfilment continues to affect suitability.
For instance, the size of a staging area yields diminishing utility and eventually saturates.
Third, the utilities of the criteria are \textit{aggregated} into a single site selection score by using additive weighted multi-criteria aggregation mechanism, and the candidate sites are ranked accordingly and mapped out spatially (Section~\ref{sec:utility}).

\subsection{Spatial Models}
\label{sec:spatialmodels}
Of the eight spatially assessable requirements, a distinction is made between constraints and criteria (Table~\ref{tab:spatialrequirements}).
The individual models compute the required metrics as follows.
\textbf{Danger zone (A1):} The danger zone depends on the hazard scenario considered, e.g. in a river flood scenario, this zone is determined by the flood extent, whereas in an UXO scenario a buffer zone is constructed around the incident location; any candidate site intersecting the danger zone is excluded.
\textbf{Manoeuvring space (B1):} The minimum inner width of a candidate polygon, i.e., the diameter of its largest inscribed circle, is computed and compared against a minimum width required for the deployment and manoeuvring of emergency vehicles.
\textbf{Heavy-vehicle driveability (B2):} A land surface layer classifies the ground cover of each candidate site; sites whose surface cannot support heavy vehicles (e.g., water bodies or forest) are excluded.
\textbf{Separate entry and exit (C2):} The street segments adjacent to a candidate site are grouped into distinct road clusters by name and topology; the site is viable only if at least two distinct clusters provide access.
\textbf{Area (C1):} The polygon area is computed in a local projected coordinate system to ensure accuracy.
\textbf{Travel times (A2, B3, F1):} Each candidate polygon is represented by its centroid and snapped to the road network, and shortest-path travel times are computed with the Dijkstra algorithm \citep{dijkstra1959}, using the driving speed for emergency vehicles of the respective OSM road type (see \citep{Priesmeier2026}).
For the proximity criterion (A2), the travel time is measured to the POIs within the danger zone.
The travel time to these facilities determines how quickly the forces can shuttle between the staging area and their places of deployment, and it is computed as the maximum network travel time from the candidate site to any of the POIs, so that each site is assessed by the facility that is hardest to reach from it.
For F1 and B3, the minimum travel times to the nearest fuel station and to the nearest main route are obtained by multi-source shortest-path searches from the respective locations.

\begin{table}[htbp]
  \centering
  \caption{Spatial models for the selected requirements. Constraints act as binary filters; Criteria are mapped to utility functions (Section~\ref{sec:utility}).}
  \label{tab:spatialrequirements}
  \small
  \setlength{\tabcolsep}{4pt}
  \renewcommand{\arraystretch}{1.15}
  \begin{tabularx}{\textwidth}{@{}l >{\raggedright\arraybackslash}p{3.2cm} >{\raggedright\arraybackslash}p{2.3cm} >{\raggedright\arraybackslash}X >{\raggedright\arraybackslash}p{3.1cm}@{}}
    \toprule
    \textbf{Req.} & \textbf{Requirement} & \textbf{Type} & \textbf{Spatial metric} & \textbf{Limit} \\
    \midrule
    A1 & Location outside danger zone & Constraint & Intersection with danger zone & zone $=\, \emptyset$ \\
    A2 & Proximity to deployment & Criterion (min) & Network travel time to the most remote POI in the evacuation area & $t \le t_{\max}$ \\
    B1 & Manoeuvring space & Constraint & Minimum inner width & $w \ge w_{\min}$ \\
    B2 & Driveability for heavy vehicles & Constraint & Land surface classification & surface suitable for heavy vehicles \\
    B3 & Connection to main routes & Criterion (min) & Network travel time to nearest main route & $U = 0$ for $t \ge t_{\max}$ \\
    C1 & Sufficient area & Criterion (max) & Polygon area & $A \ge A_{\min}$, saturation at $A_{\mathrm{sat}}$ \\
    C2 & Separate entry and exit & Constraint & Number of distinct access roads & $\ge 2$ \\
    F1 & Proximity to fuel station & Criterion (min) & Network travel time to nearest fuel station & $U = 0$ for $t \ge t_{\max}$ \\
    \bottomrule
  \end{tabularx}
\end{table}

\subsection{Utility Functions and Aggregation}
\label{sec:utility}
To compare the four criteria A2, B3, C1, and F1, and analyse how important a deviation from an optimal state is, for all criteria $c_i$ with values in $X_i$, a dimensionless utility function $U_i: X_i \to [0,1]$ is used to transform any value $x \in X_i$ to the corresponding utility $U_i (X_i)$.
Each utility function $U_i$ is defined by two anchors: the \emph{saturation anchor} $\alpha_i$, beyond which a further improvement of the metric yields no additional benefit ($U_i(x_i) = 1 \forall x_i \geq \alpha_i$\footnote{We here assume a monotonously increasing utility, if the values are to be minimised, then $\alpha_i$ is the lower and $\beta_i$ the upper bound.}), and a \emph{feasibility anchor} $\beta_i$, beyond which a site receives no utility for this criterion ($U_i(x_i) = 0 \forall x_i \le \beta_i $).
For any values $\beta_i < x_i < \alpha_i$, we use a logarithmic utility function to express diminishing returns and risk aversion \citep{Gerber01071998}: an additional 1{,}000\,m$^2$ matters considerably more for a site that barely exceeds the minimum area than for one that is already close to saturation, and an additional minute of travel weighs more heavily when the travel time is already close to the feasibility limit than when ample reserve remains.
A linear utility function would ignore this effect and would, moreover, depend on the range of values that happens to occur in the respective candidate set. A risk aversion parameter $k$ controls the curvature relative to the logarithmic case $k = 1$; $k > 1$ makes the function more tolerant of moderate deviations and $k < 1$ more strict.
Because all anchors are fixed independently of the candidate sites, the utility of a site does not depend on any other candidates; utilities are thus comparable across incidents, and adding or removing a candidate cannot reverse the order of the remaining ones, a potential weakness of normalisations that are relative to the candidate set.

For the area requirement (C1), where a larger area yields higher utility, the feasibility anchor is the minimum area required $A_{\min}$, and the saturation anchor is the area $A_{\mathrm{sat}}$ beyond which additional space is no longer beneficial:
\begin{equation}
U_{G}(A)=
\begin{cases}
0, & A < A_{\min}\\
1 - \left(1 - \dfrac{\ln (A / A_{\min})}{\ln (A_{\mathrm{sat}} / A_{\min})}\right)^{k}, & A_{\min} \le A < A_{\mathrm{sat}} \\[8pt]
1, & A \ge A_{\mathrm{sat}}
\end{cases}
\label{eq:utility_area}
\end{equation}
For the travel-time requirements (A2, B3, F1), where a lower travel time yields higher utility, the saturation anchor is a tolerance time $t_{\mathrm{good}}$, up to which a travel time is regarded as fully adequate, and the feasibility anchor is a maximum travel time $t_{\max}$, beyond which a site receives no utility for this criterion:
\begin{equation}
U_{D}(t)=
\begin{cases}
1, & t \le t_{\mathrm{good}} \\[2pt]
1 - \left(1 - \dfrac{\ln (1 + t_{\max} - t)}{\ln (1 + t_{\max} - t_{\mathrm{good}})}\right)^{k}, & t_{\mathrm{good}} < t < t_{\max} \\[8pt]
0, & t \ge t_{\max} \text{ or not reachable}
\end{cases}
\label{eq:utility_time}
\end{equation}
The offset of one minute ensures that the logarithm is defined at the feasibility anchor, where the utility reaches zero, and that the utility decreases continuously from one at the tolerance anchor; the first minutes beyond $t_{\mathrm{good}}$ therefore cost little utility, whereas further delays of a site that is already close to $t_{\max}$ reduce its assessment strongly. The anchor values used in the case study are given in Section~\ref{sec:csparam}, and the resulting functions are shown in Figure~\ref{fig:utility}.\\

The weights for the aggregation are derived from the mean category ranking of the survey (Table~\ref{tab:ranking}).
Since only categories A, B, C, and F contain soft requirements, the weight of each criterion is proportional to the reciprocal of the mean rank of its category, renormalised to sum to one:
\begin{equation}
w_i = \frac{1/\bar{r}_i}{\sum_{j \in \{A, B, C, F\}} 1/\bar{r}_j}
\label{eq:weights}
\end{equation}
This yields $w_{A2} = 0.432$, $w_{C1} = 0.243$, $w_{B3} = 0.164$, and $w_{F1} = 0.161$.
The weights are assigned at the category level, so that a criterion inherits the full weight of its category. 
Equation~\eqref{eq:weights} is one of several established ways of converting an ordinal category ranking into cardinal weights, and the choice deserves a brief justification.
Reciprocal weighting was preferred because it uses the mean rank directly, and therefore the full information of the ranking task, because it requires no additional elicitation step, and because it produces a moderate spread between the criteria.
The two common alternatives yield the same ordering: rank-sum weights \citep{stillwell1981}, $w_i = (n - r_i + 1) / \sum_j (n - r_j + 1)$ over the $n = 4$ relevant categories, result in $w_{A2} = 0.40$, $w_{C1} = 0.30$, $w_{B3} = 0.20$, and $w_{F1} = 0.10$, whereas rank-order centroid weights \citep{barron1996}, $w_i = \frac{1}{n}\sum_{j=i}^{n} 1/j$, lead to $w_{A2} = 0.52$, $w_{C1} = 0.27$, $w_{B3} = 0.15$, and $w_{F1} = 0.06$.
Compared with these schemes, the reciprocal weights are somewhat flatter, and they give B3 and F1 almost the same weight; this reflects the survey, in which the mean ranks of the two categories are nearly identical (4.40 and 4.48), a proximity that purely ordinal schemes cannot express.
For our case study, the differences between the three schemes are of the order of the weight perturbations of the Monte Carlo analysis (\ref{sec:sensitivity}), in which the core of the shortlist of top staging areas proves stable, and with rank-sum or rank-order centroid weights nine of the ten shortlisted sites and all seven core sites are retained, so the approach to determine the weights is not decisive for the result.

The utilities are aggregated into a single selection score by simple additive weighting (SAW) \citep{hwang1981}:
\begin{equation}
U_{\text{total}} = w_{A2}\, U_{A2} + w_{B3}\, U_{B3} + w_{C1}\, U_{C1} + w_{F1}\, U_{F1},
\label{eq:aggregation}
\end{equation}
where the functional form for the area criterion $C1$ is $U_G(A)$ and for the time-dependent criteria $A2$, $B3$, $F1$ we use $U_D(t)$.
The overall score $U_{\text{total}}$ allows us to rank every viable candidate site in descending order of $U_{\text{total}}$, and the top-$N$ candidates are presented to the decision maker.
Should fewer than ten candidates satisfy the constraints and the area and travel-time limits, these limit are relaxed until ten candidates can be evaluated, so that the system always returns a ranking; this fallback was not required in the case study.

\section{Case Study: Staging Area Selection for an Unexploded Ordnance Evacuation in Cologne, Germany}
\label{sec:casestudy}
To demonstrate the applicability of the selection model, we apply it to the evacuation of the Cologne-Deutz district in June 2025, in which unexploded ordnance was discovered (a scenario that often occurs in Germany \citep{Frey2023}) during survey works and a danger zone with a radius of 1{,}000\,m around the discovery site had to be evacuated prior to the defusing.
The scale of the operation was substantial: approximately 20{,}000 residents were affected, and the evacuation was complicated as major transport hubs were affected whose evacuation required substantial lead time, including three river bridges, and an important rail access to the main station.
The responsible authorities had to select staging areas decision under the time constraints of the defusing -- exactly the situation for which the proposed model provides an ad hoc and transparent ranking of all suitable candidate sites based on open data.
In the following, we describe the data used for the analysis (Section~\ref{sec:csdata}), the problem-specific threshold values of the spatial models and utility functions (Section~\ref{sec:csparam}), the resulting ranking of staging areas (Sections~\ref{sec:csresults} and~\ref{sec:csbeyond}), its comparison with the decision taken in the real operation (Section~\ref{sec:csvalidation}), and its robustness (Section~\ref{sec:csrobustness}).

\subsection{Data}
\label{sec:csdata}
The selection model relies exclusively on freely available geographic data.
The scenario-specific input is the location of the discovery site, from which the officially defined danger zone with a radius of 1{,}000\,m is buffered.
This area is used both to exclude candidate sites (A1) and to delimit the area within which the POIs (see Section \ref{sec:spatialmodels}) of the travel-time criterion (A2) are located.
All remaining inputs are taken from OpenStreetMap (OSM) as described in Section~\ref{sec:model}.

\subsection{Parameterisation}
\label{sec:csparam}
\begin{figure}[htbp]
  \centering
  \includegraphics[width=0.95\linewidth]{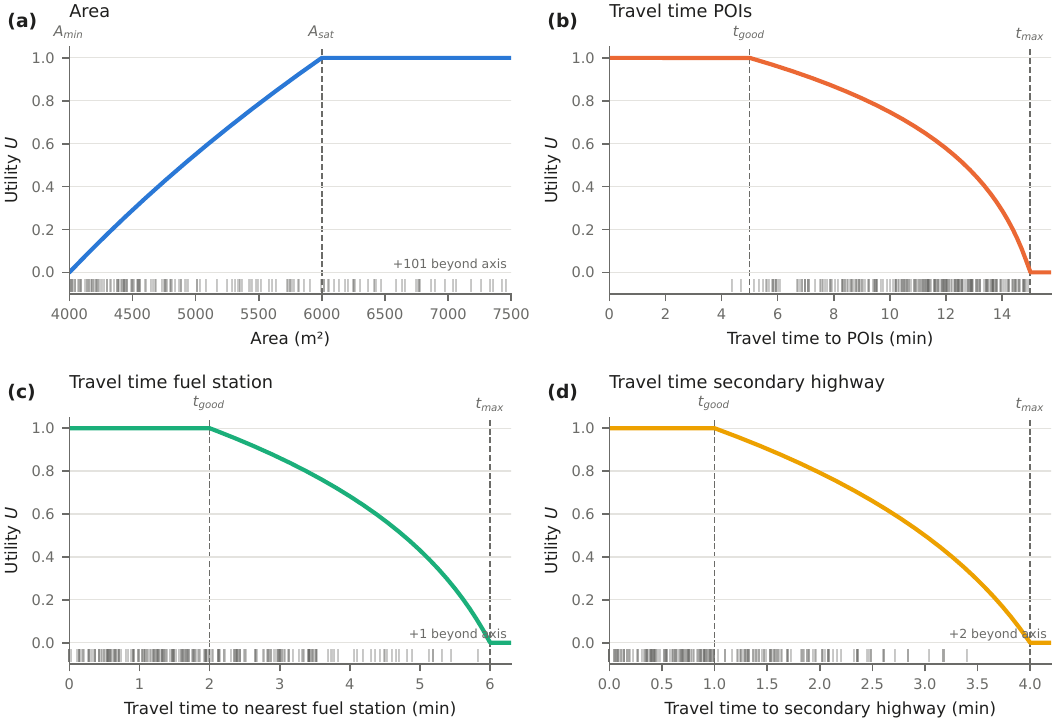}
  \caption{Utility functions of the four criteria with the anchors of the Cologne case study; the ticks below each curve mark the observed values of all evaluated candidate sites.}
  \label{fig:utility}
\end{figure}

The threshold values of the spatial models and the utility functions used in the case study are listed in Table~\ref{tab:csparameters}.
The danger-zone radius is prescribed by the responsible explosive ordnance disposal service and is therefore an exogenous scenario parameter, as is the resulting evacuation area.
The minimum required area $A_{\min} = 4{,}000$\,m$^2$ is also an exogenous input, as it derives from the number of vehicles to be accommodated: an assumed 40 vehicles requiring 50\,m$^2$ each for parking and a further 50\,m$^2$ each for lanes, manoeuvring space, and resting area. Candidate sites below this threshold are excluded.
The saturation area is set to $A_{\mathrm{sat}} = 6{,}000$\,m$^2 = 1.5\,A_{\min}$, i.e.\ a 50\,\% reserve for potentially additional units and a spatial buffer, beyond which further area is assumed to yield no additional benefit.
The minimum inner width is set to $w_{\min} = 20$\,m, so that a large emergency vehicle can turn on the site without reversing (B1).
For the travel time to the POIs, a feasibility anchor of $t_{\max} = 15$\,min is used, since a destination that cannot be reached within this time frame is no longer regarded as within operational reach, in particular for the continuous shuttle traffic between the staging area and the POIs; for A2 it simultaneously acts as a hard constraint.
For the fuel station and the main route, the feasibility anchors were derived from a city-wide reference distribution rather than from the candidate set. For every road-network node within the city boundary ($n = 16{,}408$), the travel time to the nearest target was computed, and $t_{\max}$ was set to the 99th percentile, rounded to full minutes, which yields 6\,min for the fuel station and 4\,min for the main route. These anchors thus reflect the accessibility that is realistically achievable in the study area and are independent of the discovery site and of the candidate set.
The tolerance anchors reflect the different operational role of the three destinations: $t_{\mathrm{good}} = 5$\,min for the POIs, within which a site is regarded as fully adequate for the shuttle traffic between staging area and facilities; $t_{\mathrm{good}} = 2$\,min for the nearest fuel station; and $t_{\mathrm{good}} = 1$\,min for the nearest main route, since access to the main road network should be immediate.
The latter two lie close to the medians of the reference distributions (2.5 and 0.9\,min).
Expressing service targets as thresholds of network travel time in minutes is common practice in spatial accessibility assessment \citep{hu2024accessibility}, which makes the anchors interpretable for the organisations involved even though the tolerance anchors and the limit for A2 are set by expert judgement.
The shape parameter is $k = 1$ for all functions, i.e. the utilities are purely logarithmic between their anchors.
All anchors and the shape parameter $k$ of the travel-time utilities except the exogenous minimum area are varied in the sensitivity analysis (\ref{sec:sensitivity}).
Figure~\ref{fig:utility} shows the four utility functions together with the observed values of all candidate sites.

\begin{table}[htbp]
  \centering
  \caption{Thresholds of the spatial models and anchors of the utility functions in the Cologne case study. Hard requirements act as binary filters. For the travel-time criteria, the utility is one up to $t_{\mathrm{good}}$ and zero from $t_{\max}$ onwards; for the area, sites below $A_{\min}$ are excluded and the utility is one from $A_{\mathrm{sat}}$ onwards. The shape parameter is $k = 1$ for all functions.}
  \label{tab:csparameters}
  \small
  \setlength{\tabcolsep}{4pt}
  \renewcommand{\arraystretch}{1.15}
  \begin{tabularx}{\textwidth}{@{}l >{\raggedright\arraybackslash}p{3.1cm} >{\raggedright\arraybackslash}p{2.2cm} >{\raggedright\arraybackslash}X >{\raggedright\arraybackslash}p{3.4cm}@{}}
    \toprule
    \textbf{Req.} & \textbf{Requirement} & \textbf{Type} & \textbf{Spatial metric} & \textbf{Limit} \\
    \midrule
    A1 & Location outside danger zone & Constraint & Intersection with danger zone and evacuation area & 1{,}000\,m danger zone; outside evacuation area \\
    A2 & Proximity to incident site & Criterion (min) & Max.\ network travel time to the POIs (schools, care facilities, tourist attractions, train stations) in the evacuation area & $t_{\mathrm{good}} = 5$\,min, $t_{\max} = 15$\,min (also exclusion) \\
    B1 & Manoeuvring space & Constraint & Minimum inner width & $w \ge 20$\,m \\
    B2 & Driveability for heavy vehicles & Constraint & Land surface classification & no water or forest surface \\
    B3 & Connection to main routes & Criterion (min) & Network travel time to nearest secondary road & $t_{\mathrm{good}} = 1$\,min, $t_{\max} = 4$\,min \\
    C1 & Sufficient area & Criterion (max) & Polygon area & $A_{\min} = 4{,}000$\,m$^2$ (exclusion), $A_{\mathrm{sat}} = 6{,}000$\,m$^2$ \\
    C2 & Separate entry and exit & Constraint & Number of distinct access roads & $\ge 2$ \\
    F1 & Proximity to gas station & Criterion (min) & Network travel time to nearest fuel station & $t_{\mathrm{good}} = 2$\,min, $t_{\max} = 6$\,min \\
    \bottomrule
  \end{tabularx}
\end{table}

\subsection{Results}
\label{sec:csresults}
\begin{figure}[htbp]
  \centering
  \includegraphics[width=\linewidth]{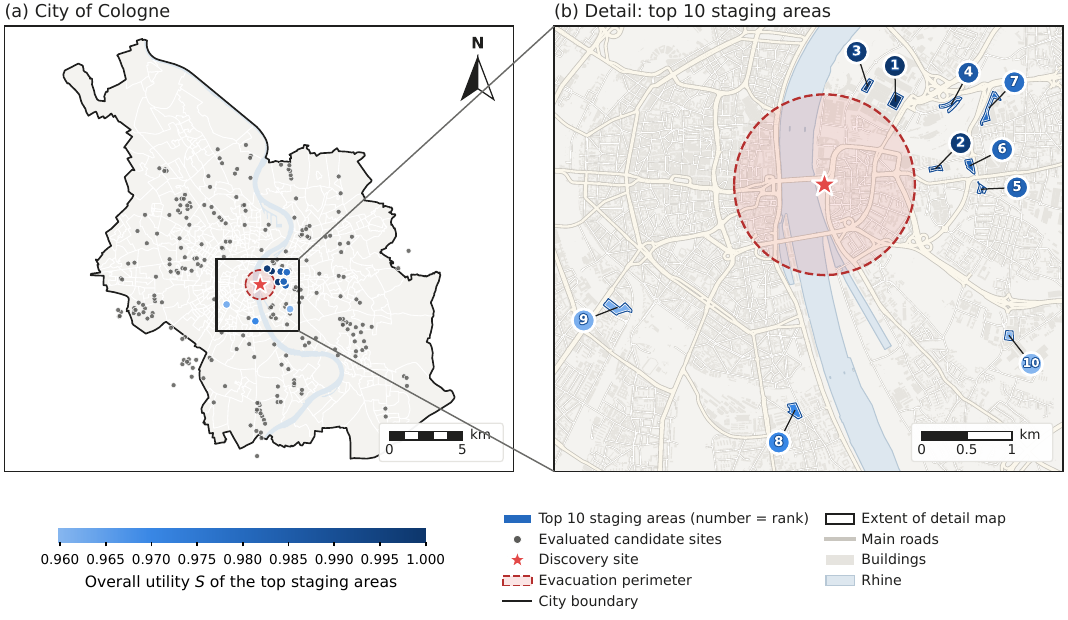}
  \caption{Map-based decision support for the Cologne case study: overview of the evacuation area with the discovery site and the ranked staging areas (left) and detailed view of the shortlisted sites (right).}
  \label{fig:exampleFig}
\end{figure}

Table~\ref{tab:csresults} lists the ten highest-ranked staging areas (the short list) together with their raw metrics and the aggregated selection value $U_{\text{total}}$; Figure~\ref{fig:exampleFig} shows the corresponding map.
Eight of the ten selected sites are located east of the Rhine and two (ranks 8 and 9) on its west bank, all in the immediate surroundings of the evacuation area, at network travel times between 4.4 and 6.7 minutes to the POIs.
The highest-ranked site is the parking area P10 with 18{,}384\,m$^2$ and a travel time of 4.7 minutes, which achieves the maximum utility in all four criteria and a selection value of $U_{\text{total}} = 1.000$.
It is followed by the parking areas P5 (Lanxess Arena) and P1, with the Koelnmesse multi-storey car park Zoobrücke P22a at rank 4; the shortlist further contains three brownfield sites (ranks 6, 8 and 9) that are large enough to accommodate a substantial number of vehicles.
This pattern is operationally plausible as areas designed for high vehicle throughput at large events combine generous dimensions with an established connection to the main road network, which is also reflected in the sites that were used in the actual operation (Section~\ref{sec:csvalidation}).\\

Three observations characterise the ranking.
First, the selection values of the top ten lie within a narrow band between 0.960 and 1.000, so the leading sites are close substitutes.
The ranking should therefore be read as a shortlist of operationally equivalent alternatives rather than as a strict rank order.
Second, the differentiation within the shortlist is driven mainly by the proximity criterion A2, whose utilities range from 0.929 to 1.000, and to a similar extent by the fuel-station criterion (F1, 0.832 to 1.000), whereas the area criterion (C1) is saturated for all ten sites and the main-route criterion (B3) for seven of them; where B3 is not saturated, its utility remains above 0.94.
The importance of A2 ($w_{A2}=0.432$) reinforces this effect, but it is the saturation of the area and main-route criteria, not the weighting alone, that produces it.
Third, the area criterion is saturated even for the smallest site (rank 5, 6{,}246\,m$^2$, $U_{C1}=1.000$), which is, consistently, the member of the group that reacts most strongly to the saturation area $A_{\mathrm{sat}}$ in the sensitivity analysis; together with the sites at ranks 9 and 10, whose selection values differ from the first challengers by less than 0.005, it forms the least stable part of the group.
Operationally, the shortlist thus consists of a robust core of seven sites and three positions for which the first challengers outside the group are equivalent alternatives.\\

\begin{table}[htbp]
  \centering
  \caption{Top ten staging areas of the Cologne case study, ranked by the aggregated selection value $U_{\text{total}}$. $A$ is the polygon area, $t_{\mathrm{POI}}$ the network travel time to the POIs in the evacuation area (A2), $t_{\mathrm{fuel}}$ and $t_{\mathrm{main}}$ the travel times to the nearest fuel station (F1) and main route (B3). $P_{10}$ the probability of top-10 membership in the Monte Carlo analysis (\ref{sec:sensitivity}).}
  \label{tab:csresults}
  \footnotesize
  \setlength{\tabcolsep}{3.5pt}
  \begin{tabular}{r l r r r r r r}
    \toprule
    \textbf{Rank} & \textbf{Site} & \textbf{$A$ [m$^2$]} & \textbf{$t_{\mathrm{POI}}$} & \textbf{$t_{\mathrm{fuel}}$} & \textbf{$t_{\mathrm{main}}$} & \textbf{$U_{\text{total}}$} & \textbf{$P_{10}$} \\
     &  &  & [min] & [min] & [min] &  &  \\
    \midrule
    1  & P10                                 & 18{,}384 & 4.68 & 0.45 & 0.70 & 1.000 & 1.000 \\
    2  & P5, Lanxess Arena                   &  7{,}182 & 5.14 & 1.92 & 0.00 & 0.998 & 0.894 \\
    3  & P1                                  &  9{,}363 & 4.37 & 2.13 & 0.84 & 0.997 & 0.999 \\
    4  & Car park Zoobr\"ucke P22a           & 15{,}570 & 5.48 & 0.58 & 1.22 & 0.985 & 0.998 \\
    5  & Unnamed parking area (b)            &  6{,}246 & 5.82 & 2.20 & 0.42 & 0.982 & 0.468 \\
    6  & Brownfield site (a)                 & 11{,}049 & 5.62 & 2.39 & 0.48 & 0.981 & 0.944 \\
    7  & Unnamed parking area (a)            & 19{,}509 & 6.07 & 1.78 & 1.10 & 0.978 & 0.994 \\
    8  & Brownfield site (b)                 & 16{,}414 & 6.69 & 1.26 & 0.64 & 0.970 & 0.798 \\
    9  & Brownfield site (c)                 & 26{,}039 & 5.69 & 3.19 & 0.92 & 0.961 & 0.554 \\
    10 & Unnamed parking area (c)            &  9{,}560 & 6.72 & 0.90 & 1.31 & 0.960 & 0.449 \\
    \bottomrule
  \end{tabular}
\end{table}

\subsection{Results beyond the shortlist}
\label{sec:csbeyond}
\begin{figure}[htbp]
  \centering
  \includegraphics[width=\linewidth]{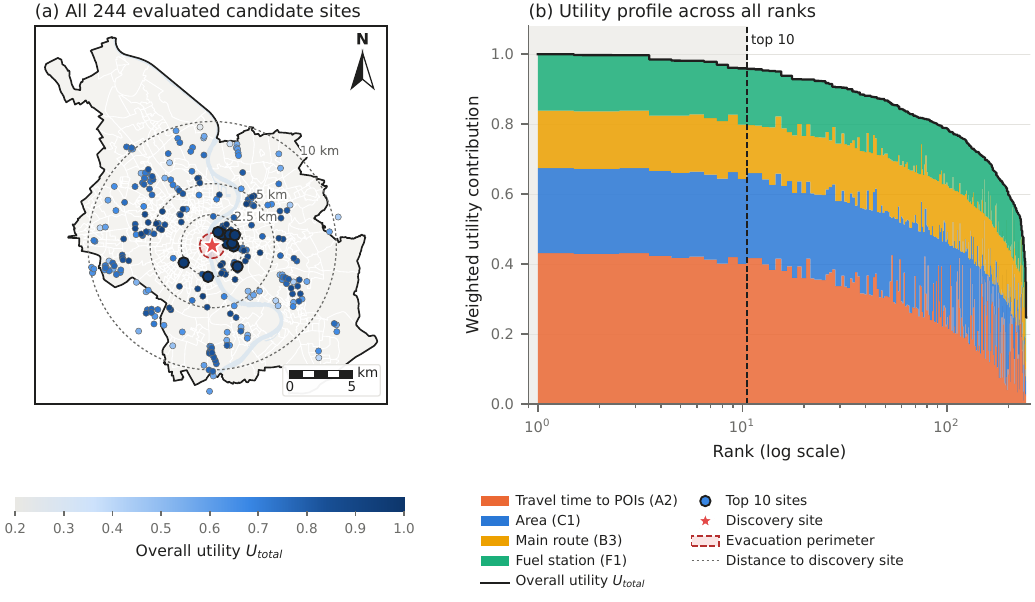}
  \caption{Results beyond the shortlist: (a) all evaluated candidate sites in the city of Cologne, coloured by the overall utility $U_{\text{total}}$, with the top ten highlighted and distance rings around the discovery site; (b) weighted contributions $w_i U_i$ of the four criteria and overall utility across all ranks (logarithmic rank axis), with the top ten shaded.}
  \label{fig:allcandidates}
\end{figure}

Figure~\ref{fig:allcandidates} places the shortlist in the context of all 244 evaluated candidate sites.
The shortlist forms the upper end of a continuum rather than a distinct group: 89 candidates reach $U_{\text{total}} \ge 0.8$, and beyond the shortlist the selection value decreases only gradually, from 0.960 at rank~10 to 0.928 at rank~20 and 0.868 at rank~50, with a median of 0.757 over all candidates.
The spatial pattern is pronounced.
The selection value decreases with the distance to the discovery site (Spearman correlation between rank and distance $\rho = 0.59$); all ten shortlisted sites lie within 2.7\,km of the discovery site (median 1.7\,km), whereas the median distance of the remaining candidates is 7.1\,km.
Proximity alone does not determine the ranking, however, of the 24 candidates within 2.5\,km, 6 remain below $U_{\text{total}} = 0.8$, since the network travel time to the POIs and the other criteria, not the straight-line distance, decide the assessment.\\

The utility profile in Figure~\ref{fig:allcandidates}(b) shows what separates the shortlist from the remaining sites.
The decisive criterion is the travel time to the POIs with its mean weighted contribution falls from 0.419 within the top ten to 0.257 for the remaining sites, reflecting travel times of 4.4 to 6.7\,minutes compared with a median of 11.6\,minutes.
The area criterion contributes a smaller part of the difference (0.243 versus 0.174), as only 53\,\% of the remaining sites reach the saturation area, compared with all ten shortlisted sites.
The fuel-station and main-route criteria barely discriminate (0.157 versus 0.147 and 0.162 versus 0.151), because most candidates in the urban area are close to both.
Below the shortlist, a second tier of 79 sites with $U_{\text{total}} \ge 0.8$ offers the natural reserve should the shortlisted sites be unavailable or should additional areas be required, and the utility profile makes explicit for which reason each of them ranks lower.

\subsection{Validation against the actual operation}
\label{sec:csvalidation}
Because the case study reproduces a real incident, the model ranking can be compared with the decision that the responsible authorities actually took.
The staging area used in the operation was the brownfield site that the model ranks sixth ($U_{\text{total}} = 0.981$).
The area actually used by the forces is thus contained in the top six of a ranking that was produced without any local knowledge, from open data and from criterion weights elicited in a survey.
This correspondence is a strong first indication of the practical validity of the approach.
The model reproduces, in a transparent and reproducible way, a choice that experienced commanders made on the basis of their familiarity with the district and of years of operational practice.
It also illustrates why the shortlist, and not the single best-ranked site, is the appropriate form of the output.
The selection values of ranks one to six differ by about 0.019, which places the site used in practice within the band of operationally equivalent alternatives.
The model narrows the search space from the full set of urban open areas to a handful of candidates, while the final allocation of roles among them -- staging area, additional parking, reserve -- remains with the commander, who can account for the factors the model does not represent.
For a municipality without comparable local expertise, or for a commander deployed outside their own district, exactly this narrowing is the decisive contribution.\\

A second reason for reporting a shortlist rather than a single recommendation is availability.
Several of the shortlisted areas are privately operated, e.g., event parking areas, and their usability depends on conditions that no static geographic data set represents, e.g., an event taking place on the day of the defusing, an area still occupied by parked cars at the time of the alert, a closed barrier, construction work, or an operator who cannot be reached within the available time.
Even where the authorities are entitled to claim such areas, access has to be arranged with the operator in practice, and this takes time.
Because the selection values within the shortlist differ only marginally, a site that turns out to be unusable can be replaced by the next candidate without a relevant loss of suitability and without re-running the analysis.
The model therefore does not answer the question which area \emph{is} the staging area, but which areas are worth to consider.

\subsection{Robustness of the ranking}
\label{sec:csrobustness}
Because both the criterion weights and the utility thresholds are estimates, the stability of the shortlist was examined by a one-at-a-time and a Monte Carlo sensitivity analysis, which are reported in detail in \ref{sec:sensitivity}.
Across all Monte Carlo runs, the perturbed top ten share on average 81\,\% of their members with the baseline shortlist, and in 3\,\% of the runs the set is identical.
The three leading sites remain within the top nine in all one-at-a-time variations, seven of the ten sites remain in the shortlist with a probability of at least 0.79, and mainly the smallest site (rank 5, $P_{10} = 0.47$) reacts noticeably to the saturation area $A_{\mathrm{sat}}$; sites from outside the shortlist enter it in up to 73\,\% of the runs (rank 11).
The median Monte Carlo ranks in Table~\ref{tab:csresults} deviate from the baseline ranks by at most one position, except for the smallest site (seven positions), which corresponds to the narrow spread of the selection values discussed above.
The shortlist should therefore be interpreted as a robust core of seven sites ($P_{10} \ge 0.79$), four of which are selected in virtually all runs, complemented by interchangeable alternatives at ranks 9--12 whose selection values differ by less than 0.005.

\section{Discussion} \label{sec:discussion}
\subsection{Principal findings}
The case study shows that the requirements elicited from practitioners can be translated into a spatially explicit model that produces a defensible shortlist of staging areas automatically and from open data.
Three findings are worth emphasising.
First, the selection can be made \emph{ad hoc and without prior knowledge of the terrain}: the candidate set is generated from open data for an arbitrary hazard location, so no pre-planned catalogue of staging areas is required, and the model can be applied in municipalities that have never prepared for such a hazard event.
Second, the result is \emph{transparent and reproducible}: identical inputs yield an identical ranking, and every rank can be traced back to the criteria utilities and their survey-based weights, which makes the rationale explainable to all organisations involved and auditable after the operation.
Third, the preferences encoded in the model do not represent a single organisation but the \emph{shared view of several stakeholder groups}, which matters because staging areas are operated jointly by fire brigades, technical relief, aid organisations, and administrations.
Beyond these properties, the case study provides a first empirical indication that the model is operationally sound as the staging area actually used in the Cologne operation appears as rank six of the ranking, although the model had no access to local knowledge beyond the geographic data from OSM (Section~\ref{sec:csvalidation}).

A methodological finding of the case study is that saturating utility functions can reduce the effective dimensionality of the decision in dense urban settings.
Once feasibility is established, in our case, most criteria are fulfilled at their plateau, see Figure \ref{fig:utility}. The ranking is governed by two criteria that discriminate -- here, the travel time to the POIs in the evacuation area and the available area -- complemented within the shortlist by the proximity to a fuel station.
This is a desirable property rather than a defect, as it mirrors operational reasoning, but it implies that the placement of the saturation and feasibility anchors deserves at least as much attention as the weights when the model is transferred to other cities or hazard types.

The situation changes as soon as the candidate set becomes more heterogeneous, which is the rule rather than the exception in other settings: with a larger danger radius the nearby large areas are excluded and the area criterion begins to bind. In suburban or rural municipalities the distances to fuel stations and main routes vary by an order of magnitude and cease to be saturated and for a larger operation, which raises $A_{\min}$, the trade-off between area and proximity becomes the decisive one.
The compensatory aggregation is therefore not redundant, but its effect is scenario-dependent, and should thus be further inspected within future research.

\subsection{Implications for practice}
For practitioners, the system offers three concrete benefits.
It shortens the time from the report of a hazard to a defensible site selection decision, since the shortlist is generated automatically as soon as the evacuation parameter is known (e.g. in an UXO scenario) or the hazard extent changes (e.g. in a dynamic flood scenario).
It makes the decision communicable using the ranked map with explicit criteria that support the coordination between the organisations involved and the justification of the choice towards the political level and the public.
And it provides an overview of possible fallbacks. If the top candidates are close substitutes in terms of their selection score, the commander can switch to the next site without repeating the analysis if a site turns out to be occupied, blocked, hosting an event, or otherwise unavailable -- a case that can be expected with the privately operated areas that dominate the shortlist of the case study (Section~\ref{sec:csvalidation}). If the differences among candidates are greater, the ranking can support a deliberation about the next best and viable alternatives.

\subsection{Implications for research}
The work points towards three intriguing research directions.
The first concerns the decision problem itself: large evacuations frequently require \emph{several} staging areas, for different organisations or for different sectors of the evacuation area, which turns the present ranking problem into a covering or assignment problem with interaction effects between the selected sites.
The second concerns uncertainty: travel times derived from speed limits ignore congestion, road closures caused by the evacuation itself, and the temporal evolution of the operation, so that stochastic or time-dependent network models would be a natural extension.
Network reliability models, which treat rescue travel times as stochastic and account for disruptions such as road damage and high traffic density, offer formulations of this extension \citep{Priesmeier2026,huang2024rescuenetwork}; parts of this approach are already included in the present model.

Beyond a single incident as presented in the case study, the model can be used ex ante as a planning instrument, by running it for a multitude of hypothetical danger/hazard zones thereby identifying the areas that are repeatedly selected across scenarios, as these areas are natural candidates for preparatory agreements with their owners.
In the context of UXO sites, the spatial arrangement of UXO locations could e.g. be informed by the methodology of Mahling et al. \citep{Mahling2012}, who proposed a statistical approach for identifying patterns in UXO distributions.

Finally, the transfer of the approach to other ad-hoc siting problems -- collection points or helicopter landing areas -- would test whether the combination of survey-based weights, hard filters, and saturating utilities generalises beyond the staging area context.

\subsection{Limitations}
Several limitations of the study should be noted, concerning the survey, the model, and the data.
First, the survey was distributed via the mailing lists and individual distributors of a limited number of organisations and relied on self-selection; hence the respondents do not constitute a representative sample, and the results reflect the perceptions of a self-selected group of German practitioners. Applying our approach in other contexts will require a careful consideration of the resulting ranking and weights. 
Second, the Likert ratings exhibit a ceiling effect, as nearly all requirements are rated as important or very important; the per-requirement ratings therefore provide limited discriminative power, and the category ranking should be regarded as the primary basis of the prioritisation. 

With respect to the model, the restriction to spatially assessable criteria means that a site which the model ranks highly may still be rejected for reasons outside the model, such as ownership, contractual availability, ongoing events, or the presence of underground structures that limit the axle load; the system therefore supports, but does not replace, the judgement of a commander.
The weights are inherited at the category level, so that a single soft criterion can carry the entire weight of its category, and the utility thresholds -- in particular the saturation area and the maximum travel times -- are expert estimates rather than empirically derived quantities.

Finally, the data quality of OSM is decisive for the result \citep{haklay2010}, e.g. for the completeness of open areas, the correctness of land-use tags, and the presence of speed limits vary between regions.
The correspondence between the model ranking and the sites actually used in the Cologne operation (Section~\ref{sec:csvalidation}) is encouraging, but it rests on a single incident; systematic validation across several operations and exercises, and against the decisions actually taken by incident commanders, remains the most important next step.

\section{Conclusion}
This paper presents a spatial decision support system for the ad hoc selection of emergency staging areas using open data.
Its contribution is twofold: a set of requirements and relative priorities elicited from 104 civil protection practitioners, and a computational model that translates the spatially assessable subset of these requirements into hard feasibility filters and saturating utility functions over candidate areas derived from open data, aggregated into a transparent ranking.
Applied to a real evacuation in Cologne, the system produced from freely and openly available data a shortlist of ten staging areas that proved robust against plausible variations of the weights and thresholds and that contains the actual staging area used in the operation among its six leading candidates.
The approach thereby replaces an implicit, experience-based judgement by an explicit and reproducible one, while keeping the commander in control -- and it offers a template for the many other ad-hoc siting decisions that emergency services have to take under time pressure.

\section*{Declaration of generative AI and AI-assisted technologies in the manuscript preparation process}
During the preparation of this work the authors used Claude (Anthropic) to support code development, to search for selected references and to support the editing of selected text passages. After using this tool, the authors reviewed and edited the content as needed and take full responsibility for the content of the published article.

\appendix
\section{Appendix}
\subsection{Sensitivity analysis}
\label{sec:sensitivity}
The robustness of the baseline top~10 was tested with respect to the survey weights and the anchors of the utility functions; $A_{\min}$ was kept fixed as an exogenous requirement. First, each parameter was varied individually over 31 equidistant values within a plausible interval (baseline in brackets): $t_{\mathrm{good}}$ for the POIs 3--6~min (5), for fuel stations 1--3~min (2), and for main routes 0.5--2~min (1); $t_{\max}$ for the POIs 12--18~min (15), for fuel stations 5--8~min (6), and for main routes 3--6~min (4); $A_{\mathrm{sat}}$ 5{,}000--8{,}000~m\textsuperscript{2} (6{,}000); and the shape parameter $k$ of the travel-time utilities 0.7--1.5 (1). Second, all inputs were perturbed simultaneously in 5{,}000 Monte Carlo runs \citep{butler1997}, with weights drawn from a Dirichlet distribution centred on the survey weights (concentration 50) and parameters drawn uniformly from the same intervals.

\begin{figure}[htbp]
  \centering
  \includegraphics[width=\linewidth]{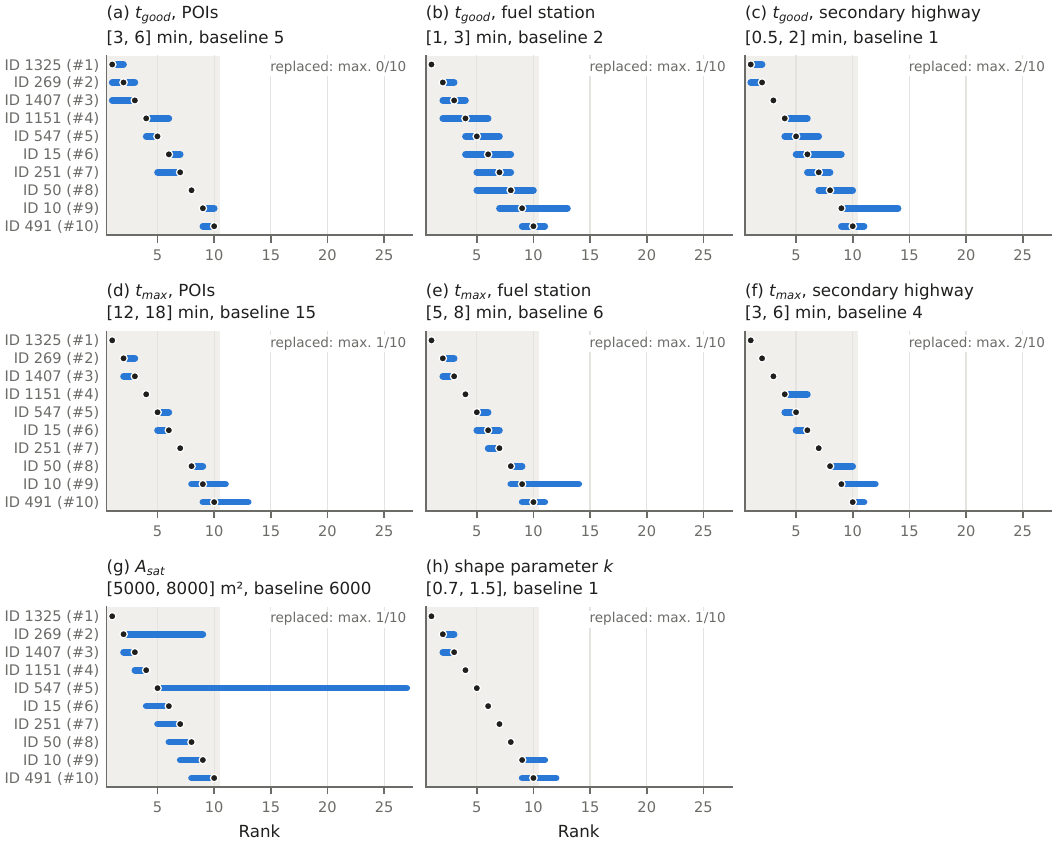}
  \caption{One-at-a-time sensitivity of the baseline top~10.}
  \label{fig:sens_oat}
\end{figure}

\begin{figure}[htbp]
  \centering
  \includegraphics[width=\linewidth]{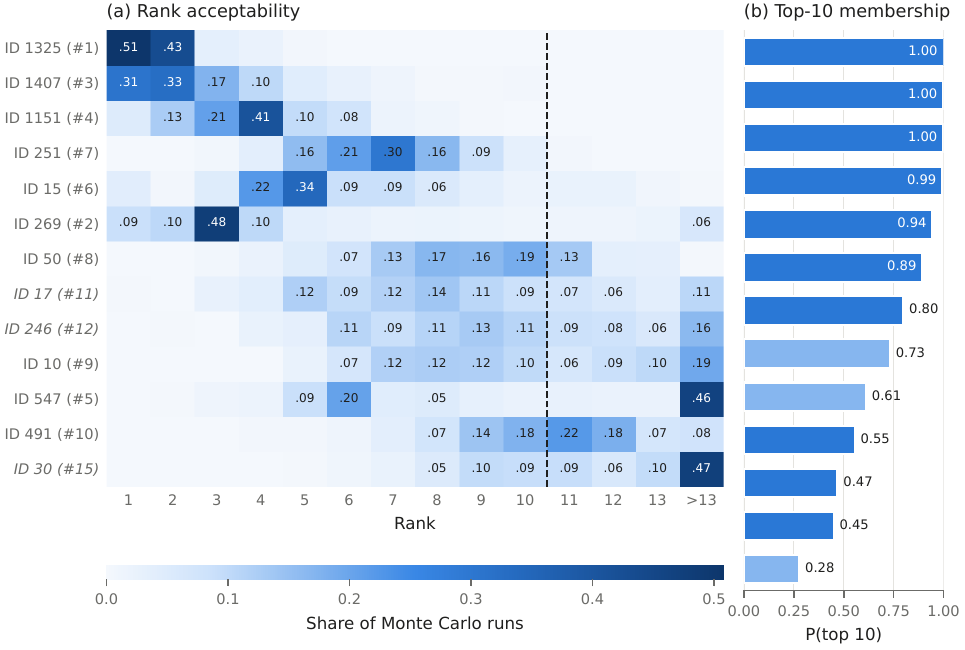}
  \caption{Monte Carlo sensitivity: (a)~rank acceptability \citep{lahdelma1998}, (b)~probability of top-10 membership; sites outside the baseline top~10 in italics.}
  \label{fig:sens_mc}
\end{figure}

\begin{figure}[tbp]
  \centering
  \includegraphics[width=0.72\linewidth]{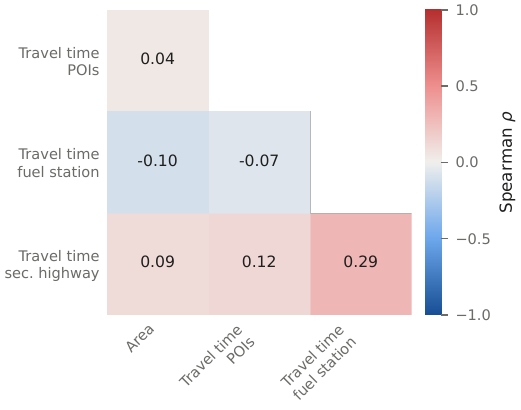}
  \caption{Spearman rank correlation between the evaluation criteria.}
  \label{fig:corr}
\end{figure}

In the one-at-a-time analysis (Figure~\ref{fig:sens_oat}; bars show the range of ranks, dots the baseline rank), $k$ has virtually no effect (maximum rank shift of two positions) and $t_{\max}$ a moderate one (maximum rank shift of five positions), and the three leading sites remain within the top nine across all intervals. At most two of the ten sites are replaced, and only when $t_{\mathrm{good}}$ or $t_{\max}$ of the main routes is varied, in which case the sites ranked eleventh, twelfth or fifteenth enter the group; the remaining parameters replace at most one site, and only $t_{\mathrm{good}}$ of the POIs leaves the composition of the top ten unchanged. Reorderings within the group are driven mainly by the anchors of the fuel-station and main-route utilities and by $A_{\mathrm{sat}}$. The Monte Carlo analysis (Figure~\ref{fig:sens_mc}) confirms a stable core but shows that the margin of the group is interchangeable: the perturbed top ten overlap with the baseline group by 81\,\% on average (5th percentile 70\,\%) and are identical in 3\,\% of the runs; four sites (ranks 1, 3, 4 and 7) are selected in virtually every run ($P(\mathrm{top~10}) \geq 0.99$), and three further sites of the baseline group are retained with high probability ($\geq 0.79$), whereas the sites ranked fifth, ninth and tenth are retained in only 45--55\,\% of the runs and challengers from outside the group enter it frequently (up to 73\,\% of the runs, for the site ranked eleventh). Mainly the smallest site (rank 5; 90\,\% rank interval 5--44) responds noticeably to the choice of $A_{\mathrm{sat}}$. Furthermore, the four criteria are weakly correlated (Figure~\ref{fig:corr}; Spearman $|\rho| \leq 0.29$), so each contributes independent information and no aspect is counted twice.

Overall, the selection of the top staging areas is robust with respect to its core, as seven sites remain in the top ten in at least 79\,\% of the Monte Carlo runs and four of them in virtually all runs, whereas ranks 9--12 are separated by score differences below 0.005 and should be regarded as interchangeable alternatives. The leading sites are identified consistently, which, complemented by a short list of equivalent alternative sites, provides a reliable basis for operational planning within the parameter ranges examined.

\bibliographystyle{elsarticle-num}
\bibliography{references}

\end{document}